# The Perception of Humanoid Robots for Domestic Use in Saudi Arabia

**Ohoud Alharbi**

Simon Fraser University

Vancouver, BC, Canada

oalharbi@sfu.ca

**Ahmed Sabbir Arif**

University of California, Merced

Merced, CA, USA

asarif@ucmerced.edu

**Abstract**

We propose a research to investigate Saudi peoples' perception of humanoid domestic robots and attitude towards the possibility of having one in their house. Through a series of questionnaires, semi-structured interviews, focus groups, and participatory design sessions, this research will explore Saudi peoples' level of acceptance towards domestic robots, the tasks and responsibilities they would feel comfortable assigning to these robots, their preferred appearance of domestic robots, and the cultural stereotypes they feel a domestic robot must mimic.

**ACM Classification Keywords**

I.2.9. Robotics: Commercial robots and applications; H.1.2. User/Machine Systems: Human factors.

**Introduction**

In 2017, Saudi Arabia became the first country in the world to give a humanoid robot named Sophia citizenship [20]. Sophia is created by a Hong Kong based robotics company, Hanson Robotics. She speaks only English and does not look like a person from the

Arab region (Figure 1). Since Saudi Arabia has a strict citizenship policy and is a deeply religious, conservative, traditional, and family-oriented society, granting Sophia citizenship raises many interesting sociotechnical questions. However, this paper focuses *only* on an upcoming research that will investigate Saudi peoples' perception about humanoid domestic robots. We find this topic particularly interesting in light of the facts that: domestic workers are more common in the Middle East [7] than the other parts of the world; employing domestic workers has and has become a symbol of social status in the Middle East [7]; and but also about 99.6% of domestic workers in Saudi Arabia are non-native [18].

## Related Word

Some researchers have investigated the sociocultural aspects of different types of robots. This section discusses the most relevant works in the area.

*Attitude Towards Robots*
In the past, many have studied peoples' attitude towards robots. Friedman et al. [5] and Kahn et al. [8] evaluated AIBO, a robot designed to mimic dogs' behavior and appearance (Figure 2) [13]. Their research involved unstructured playing sessions with children and online discussion forums with adults. Results showed that both adults and children found AIBO engaging.

Khan [10] investigated adults' attitude towards robots using a questionnaire that addressed peoples' thoughts about domestic robots' appearance and behaviors. The results revealed that most participants were positive about the idea of having a robot at home. Scopelliti et al. [17], in contrast, studied peoples' perception of domestic robots. In their study, they recruited participants from three different generations: young adults (18-25 years), middle aged (40-50 years), and elderly (65-75 years). They found out that young adults have a more positive attitude towards domestic robots than the older age groups.

*Robots Replacing Humans*
Although humanoid robots have the potential to become useful assistants for society following the tradition of automobiles and personal computers [9], important issues involving human-robot interaction, starting from physical touch to gestures and spoken languages, need to be addressed. Humanoid robots must make their users and other humans feel comfortable around them and fit in with daily life [9]. Pransky [14] discussed the weaknesses of replacing humans with robot in jobs that need social interaction. For instance, he argued that robots replacing nannies could prevent children from having a normal human interaction and could create perception that robot interaction is the norm.

*Effects of Cultural Norms*
Robots are rooted in our cultural expectations as *"servant, enemy, friend, pet, slave, toy, companion, and other roles presented in popular mythology"* [11]. These roles frame user stereotypes [2]. Thomas [19] and Rogers [16] emphasized on the importance of understanding the social and cultural norms of a country to facilitate technology acceptance by its people. Riek et al. [15] conducted a study to understand Arab peoples' views of owning humanoid robots. The results indicated that their attitude toward humanoid robots is mostly positive. Interestingly, participants from the Gulf region (Saudi Arabia, Iran, Iraq, Oman, Qatar,

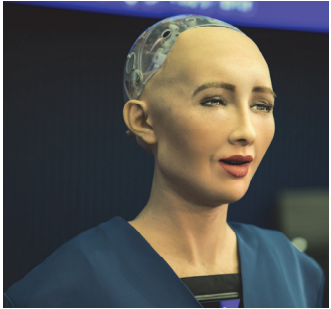

**Figure 1.** Humanoid robot Sophia spoke at the AI for Good Global Summit 2017 in Geneva, Switzerland. From ITU Pictures https://flic.kr/p/UivkB3.

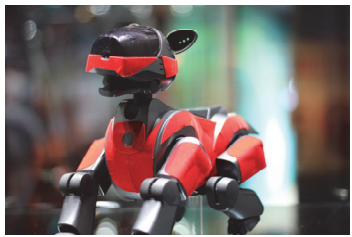

**Figure 2.** Robot dog AIBO. From Sven Volkens.

United Arab Emirates, and Yemen) had significantly more favorable views toward humanoid robots than the ones from the African region (Egypt, Morocco, Tunisia, Libya, and Sudan). This indicates towards the possibility that cultural attitudes affect the acceptance of humanoid robots. In our research, we will attempt to understand the variety of cultural attitudes in Saudi households and their effects on the acceptance of humanoid robots.

*Robot Appearance*
A robot's appearance can shape the social expectations and can impact peoples' perceptions in terms of likeability, believability, and engagement with the robot [4]. A robot with animal appearance is likely to be interpreted differently than a robot with human-like appearance [6]. In a prior study [12], participants frequently comment on a robot's perceived gender, race or nationality, and social standing within the household. This suggests that our natural tendency to categorize others persist even with humanoid robots.

## Research Questions

This exploratory research will investigate Saudi peoples' perception of humanoid robots and their attitude towards the possibility of having one in their house. More specifically, it will attempt to answer the following questions through various questionnaires, semi-structured interviews, focus groups, and participatory design sessions.

1) *Do* Saudi people accept the idea of *having a domestic robot in their house?*

2) *What are Saudi peoples' perceptions about domestic robots?*

3) *What tasks and responsibilities do Saudi people want domestic robots to perform?*

4) *What are Saudi peoples' preferred appearance of domestic robots?*

5) *Which cultural norms and stereotypes they think domestic robots must mimic?*

## Participants

For this research, we will recruit Saudi middle-high (income above 22,900 SAR [1]) to upper-class (income above 38,200 SAR [1]) families who employ domestic workers. The families must have a minimum household size of two. All household members, regardless of their age and gender, will be recruited.

With the help of local collaborators, we will distribute invitations to participate in our research at various online mailing lists, forums, newspapers, and magazines. We will also set up recruitment booths at public places, such as shopping malls. All participants will be compensated for their time with gift cards or complementary meals at a popular restaurant.

## Design

This research will use a mixed method since it is beneficial in investigating new perspectives. This will enable us to obtain divergent information by modifying the questions of the follow-up methods based on the results of the previous method(s).

*Interviews and Focus Groups*
This research will start with several semi-structured interviews and focus groups with each participating family. During the interviews, families will be asked about their opinions of domestic robots. During the focus groups, multiple videos of humanoid robot will be displayed to encourage discussions about different aspects of humanoid robots. All responses will be recorded through observation and interview notes.

*Participatory Design*
There will be a participatory design session focusing on robot appearance following the interviews and focus groups. This session will enable us to elaborate, illustrate, and clarify the accepted appearance of humanoid robots. Furthermore, the findings of this session can help us prepare a set of more fine-tuned questions for the follow-up sessions. For example, the designs/sketches of robot appearance collected from this session can help us lead a better discussion by providing more references on robot appearance.

*Questionnaires*
This research will collect two questionnaires, one before and another after the interviews and focus groups.

The pre-interview/focus group questionnaire will be an adaption of the Cogniron introductory questionnaire [3] that will ask participants about their personal details, including gender, age, occupation, as well as their level of familiarity with robots, prior experience with robots, and their level of technical knowledge of robots. The post--interview/focus group questionnaire will be an adaption of the Cogniron final questionnaire [3] that will ask participants about their perceptions about robot appearance, robot roles, robot behavior, and robot communication.

This research will conduct both interviews and focus groups and questionnaires in an attempt to collect a comprehensive dataset. Questionnaires will provide us with an insight into how users perceive humanoid domestic robots. Interviews and focus groups, on the other hand, will help us in identifying the source and implications of these perceptions by detecting factors that might be otherwise missed. In addition, the results from different methods could validate each other, providing stronger evidence for a conclusion.

## Ethics
This research will follow the rules governing the ethics of scientific research issued by King Saud University's council. Before conducting the studies, we will get an approval from King Saud University (KSU) in Riyadh, Saudi Arabia in collaboration with the Human-Computer Interaction (HCI) Lab at KSU.

All Participants will be treated equitably and fairly. Participants will be provided with information about the research and its validity for the larger community. They will be informed about the potential risks, which for this study are minimal, and that they can withdraw from the study at any time before, during, and after the study. Participants will sign a consent form, which will also provide them with information on how their personal information will be handled and safeguarded and how the results of this research will be disseminated.

## Conclusion
This research will explore Saudi peoples' level of acceptance towards humanoid domestic robots, the types of tasks and responsibilities they would assign to these robots, their preferred appearance of domestic robots, and the cultural stereotypes they feel domestic robots must imitate. To investigate this, this research will use a mixed method composed of questionnaires, semi-structured interviews, focus groups, and participatory design sessions.

We hope the findings of this research will provide an insight into Saudi peoples' needs, desires, and expectations of humanoid domestic robots, enabling the design of more effective, useful, and socially acceptable robots. We also hope that this will inspire further research in this area.